\documentclass[twocolumn, switch]{article} 
\usepackage{preprint}

\usepackage[utf8]{inputenc} 
\usepackage[T1]{fontenc}    
\usepackage{amsmath}
\usepackage{amsfonts}
\usepackage{mathpazo}
\usepackage{inconsolata}                             
\usepackage{xcolor}
\usepackage[hyphens,obeyspaces,spaces]{url}
\usepackage{hyperref}
\usepackage{multicol}
\usepackage{xspace}                                
\usepackage[detect-all]{siunitx}
\usepackage[font=small]{caption}
\usepackage{bm}
\usepackage{microtype}      
\usepackage{float}			
\usepackage{authblk}

\graphicspath{{figures/}}                           

\newcommand{\figref}[1]{Fig.~\ref{#1}}

\newcommand{\vv}[1]{\bm{\mathrm{#1}}}                   

\def\citenum#1{\def\@cite##1##2{##1}\cite{#1}}

\makeatletter      
   \newcommand{\figcaption}{\captionsetup{type=figure} \def\@captype{figure}\caption}            
   \newcommand{\tabcaption}{\captionsetup{type=table} \def\@captype{table}\caption}            
   \renewcommand{\@biblabel}[1]{#1.}    
\makeatother

\title{Deciphering Pancharatnam's discovery of geometric phase}

\author[1]{Luis Garza-Soto}
\author[1\thanks{\tt{nh@hagenlab.org}}]{Nathan Hagen}
\author[2]{Dorilian Lopez-Mago}

\affil[1]{Department of Optical Engineering, Utsunomiya University, 7-1-2 Yoto, Utsunomiya, Tochigi 321-8585 Japan}
\affil[2]{Tecnologico de Monterrey, Escuela de Ingenier{\'\i}a y Ciencias Ave.\ Eugenio Garza Sada 2501, Monterrey, N.L., M{\'e}xico, 64849}

\begin{document}

\twocolumn[ 
  \begin{@twocolumnfalse} 
  
\maketitle

\begin{abstract}
   While Pancharatnam discovered the geometric phase in 1956, his work was not widely recognized until its endorsement by Berry in 1987, after which it received wide appreciation. However, because Pancharatnam's paper is unusually difficult to follow, his work has often been misinterpreted as referring to an evolution of states of polarization, just as Berry's work focused on a cycle of states, even though this consideration does not appear in Pancharatnam's work. We walk the reader through Pancharatnam's original derivation and show how Pancharatnam's approach connects to recent work in geometric phase. It is our hope to make this widely cited classic paper more accessible and better understood.
\end{abstract}
\vspace{0.35cm}

  \end{@twocolumnfalse} 
] 

\section{Historical introduction}

Sivaramakrishnan Pancharatnam was born in Calcutta in 1934 in a family of remarkable scientists. He joined his brother S.~Ramaseshan at Bangalore in the Raman Research Institute in 1953 when he was 19 years old.\cite{Ramaseshan1988,Nityananda2013} At that time Ramaseshan's research supervisor was C.~V.~Raman, their uncle, and who received the Nobel prize in physics in 1930. Under C.~V.~Raman's instruction, Pancharatnam studied the behavior of absorbing bi-axial crystals, as he points out in the opening of his now-famous paper, ``Generalized theory of interference and its applications, part I. Coherent pencils''.\cite{Pancharatnam1956} This impressive work was published in 1956 when Pancharatnam was 22. In it, Pancharatnam proposes a definition for two beams of different states of polarization to be in phase: ``The phase advance of one polarised beam over another (not necessarily in the same state of polarization) is the amount by which its phase must be retarded relative to the second, in order that the intensity resulting from their mutual interference may be maximum.''\cite{Pancharatnam1956} Berry named this \textit{Pancharatnam's connection}.\cite{Berry1987} 

After providing expressions for the phase shift --- what we now refer to as the geometric phase --- Pancharatnam goes on to relate the expressions to the subtended solid angle on the Poincar{\'e} sphere. In Sections~\ref{sec:start} \& \ref{sec:solid_angle} below, we retrace Pancharatnam's derivation using a more modern approach that should be easier for modern readers to follow. Along the way, it becomes clear that Pancharatnam nowhere considers the case that is widely attributed to him, that the phase of polarized light changes after a cyclic evolution of its polarization.\cite{Berry1987,Aravind1992,Tiwari1992,Roy2002,Kurzynowski2011,Lages2014,Cohen2019,Arteaga2021} This misconception of his actual work appears to be a result of the difficulty one encounters when reading Pancharatnam's paper, which often feels like it belongs to the 19th century, and also of confounding Pancharatnam's work with the closely related work of Berry.

Almost 30 years after Pancharatnam's original work, Michael Berry at the University of Bristol, unaware of Pancharatnam's work, discovered that an unexpected phase emerges after the adiabatic evolution of a quantum state around a cycle in parameter space.\cite{Berry1984} In 1983, before his initial paper was published, Berry introduced the geometric phase to Barry Simon, who immediately coined it as \textit{Berry's phase}.\cite{Simon1983} By the end of 1986 Ramaseshan and Nityananda revived Pancharatnam's work and presented it as an example of Berry's phase.\cite{Ramaseshan1986} Berry himself received Nityananda's manuscript and read it, but mentions that it was not until he visited Bangalore in July 1987 that he came to appreciate Pancharatnam's work. One can surmise that Nityananda's interpretation of Pancharatnam's phase was likely greatly influenced by Berry's personal approach to the geometric phase via cycles of states. This seems likely because Berry, on his return from India, prepared a new manuscript that revealed the connections between his and Pancharatnam's work, in which he explains Pancharatnam's work in terms of cycles of states.\cite{Berry1994,Berry2010} Subsequent researchers have almost invariably followed this interpretation, with the result that Pancharatnam's actual achievement is obscured beneath the misconception.

As Berry himself pointed out, there have been several anticipations to geometric phase that arise before Pancharatnam's work.\cite{Berry2010} Vinitskii \textit{et al.}, for example, mention the work of Rytov (1938) and Vladimirskii (1941) as precursors.\cite{Vinitskii1990} Oriol Arteaga has also pointed out that Fresnel and Arago in 1816 developed their ``fifth'' law of optical interference in such a way that a geometric phase term (at the time not well understood) had to be included in the interference equations.\cite{Arteaga2021} However, as is commonly the case in science, every discovery can be traced to its anticipations, and we focus on Pancharatnam because his achievement is widely recognized by the scientific community.\cite{Jackson2008}


\section{Pancharatnam's starting point}\label{sec:start}

\begin{figure}
    \centering
    \includegraphics[width=0.65\linewidth]{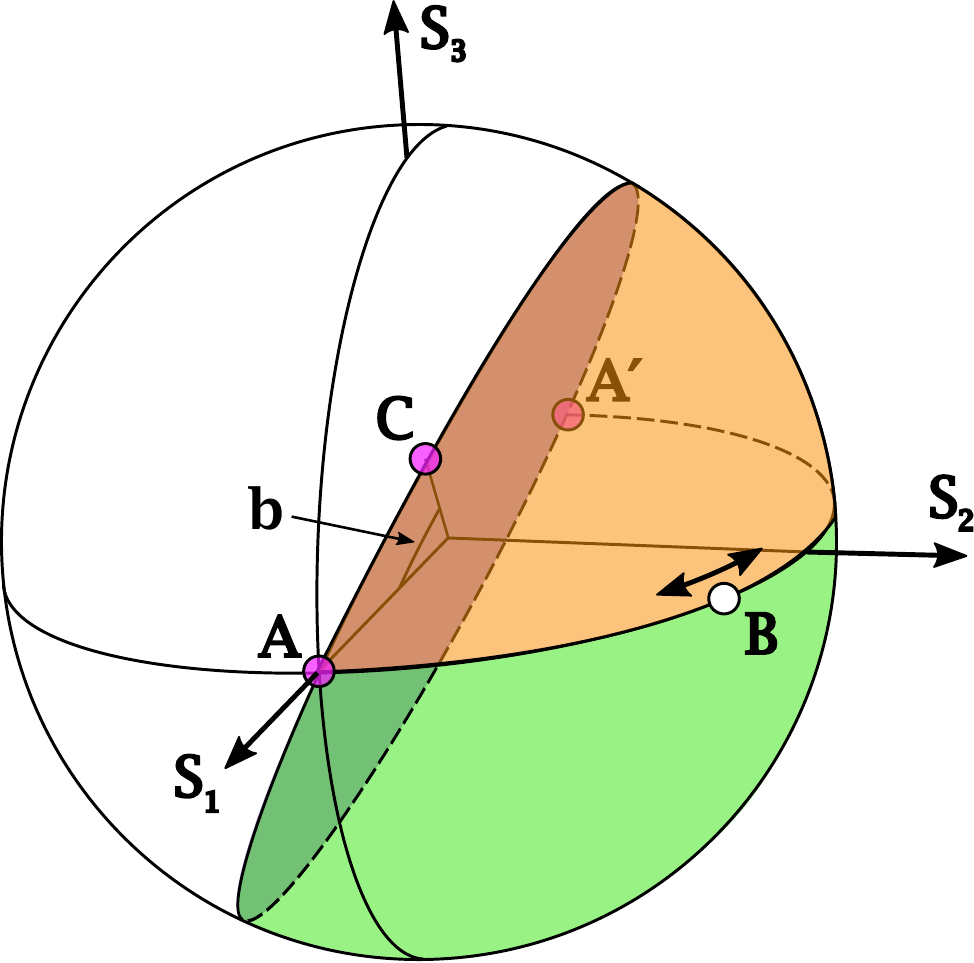}
    \caption{Polarization state $\mathbf{C}$ with respect to the two orthogonal states $\mathbf{A}$ and $\mathbf{A'}$, represented on the Poincar{\'e} sphere. Point $\mathbf{B}$ represents an SOP lying between $\mathbf{A}$ and $\mathbf{A'}$.}
    \label{fig:1}
\end{figure}

At the beginning of his manuscript, after explaining briefly some of the properties of the Poincar{\'e} sphere and Stokes parameters, Pancharatnam introduces the following theorem: 
\begin{quote}
     \textit{Theorem 1}. When [an electromagnetic] vibration of intensity $I$ in the state of polarisation $\mathbf{C}$ is decomposed into two vibrations in the opposite states of polarisation $\mathbf{A}$ and $\mathbf{A'}$, the intensities of the $\mathbf{A}$-component and the $\mathbf{A'}$-component are $I \cos^2 (AC/2)$ and $I \sin^2 (AC/2)$ respectively.
\end{quote}
(See \figref{fig:1} for an illustration of the geometry.) 
Here Pancharatnam makes use of the fact that the angle $\angle A'C$, between points $\mathbf{A'}$ and $\mathbf{C}$, subtended from the center of the Poincar{\'e} sphere, is complementary to angle $\angle AC$, and so writes both components in terms of $\angle AC$ alone. Since we will be making use of these angles in many of the equations below, we follow Pancharatnam and define the individual angles $a$, $b$, and $c$ as
\begin{equation}\label{eq:abc}
\left.
   \begin{aligned}
      c/2 &= \angle AB \\
      b/2 &= \angle AC \\
      a/2 &= \angle BC \phantom{ZZZZ}
   \end{aligned}
\right\}
\end{equation}
Note that these angles $a$, $b$, and $c$ on the left hand side of the equations are defined in terms of the electric fields in Cartesian space, whereas the arcs on the right hand side are defined in Poincar{\'e} space, hence the division by two in each definition.

Since this theorem is the basis for much of Pancharatnam's subsequent results, we show how one can derive it using a modern approach. Let us consider a monochromatic electromagnetic wave propagating along the $z$ axis, i.e., $\mathbf{E} = \mathbf{E}_C \exp[i(kz-\omega t)]$. The electric field amplitude $\mathbf{E}_C$ can be written using elliptical polarization states $\{\hat{\mathbf{e}}_A, \hat{\mathbf{e}}_{A'}\}$ as a basis, with the properties $|\hat{\mathbf{e}}_{A,A'}| = 1$, and $\hat{\mathbf{e}}_A \cdot \hat{\mathbf{e}}_{A'}^\ast = 0$ (where $\ast$ represents the complex conjugate). Therefore,
\begin{equation}\label{eq:Ec}
    \mathbf{E}_C = E \cos (\alpha) \hat{\mathbf{e}}_A + E \exp(i\beta) \sin (\alpha) \hat{\mathbf{e}}_{A'},
\end{equation}
where $E$ is a real amplitude (thus, $I = E^2$), $\alpha$ controls the projection over the basis vectors, and $\beta$ is the relative phase. With the above equation, we can represent any state of polarization (SOP) by modifying $\alpha$ and $\beta$. 

For the common definition of the Stokes parameters, we would choose $\hat{\mathbf{e}}_A = \hat{x}$ and $\hat{\mathbf{e}}_{A'} = \hat{y}$, so that
\begin{equation}
   \left.
   \begin{aligned}
      \mathbf{S}_0 &= |E_x|^2 + |E_y|^2 \, , \phantom{ZZZZ} \\
      \mathbf{S}_1 &= |E_x|^2 - |E_y|^2 \, , \\
      \mathbf{S}_2 &= 2 \mathrm{Re} [E_x E_y^\ast] \, , \\
      \mathbf{S}_3 &= 2 \mathrm{Im} [E_x E_y^\ast] \, ,
   \end{aligned}
   \right\}
\end{equation}
However, this is only one specific choice. An equally valid choice is the following general form of the Stokes parameters\cite{collett2005field}
\begin{equation}
   \left.
   \begin{aligned}
      \mathbf{S}_0 &= |E_{A}|^2 + |E_{A'}|^2 \, , \phantom{ZZZZ} \\
      \mathbf{S}_1 &= |E_{A}|^2 - |E_{A'}|^2 \, , \\
      \mathbf{S}_2 &= 2\mathrm{Re} [E_{A} E_{A'}^\ast] \, , \\
      \mathbf{S}_3 &=2\mathrm{Im} [E_{A} E_{A'}^\ast] \, ,
   \end{aligned}
   \right\}
\end{equation}
where $E_j$, with $j \equiv {A,A'}$, are the components of $\mathbf{E}_C$, i.e., $E_j = \mathbf{E}_C \cdot \hat{\mathbf{e}}_j$. Let us consider $\mathbf{C} = \{S_1,S_2,S_3\}$ as the Stokes vector representing the SOP of $\mathbf{E}_C$. $\mathbf{C}$ lies on the surface of the Poincar\'e sphere of radius $S_0$ with main axes $\{\mathbf{S}_1,\mathbf{S}_2,\mathbf{S}_3\}$. These definitions are illustrated in \figref{fig:1}, where the $\mathbf{S}_i$ are with respect to a general elliptical basis and need not be with respect to the $x$-$y$ basis. By using \eqref{eq:Ec}, the Stokes parameters for $\mathbf{E}_C$ are
\begin{equation}\label{eq:stokesvector_alphabeta}
   \left.
   \begin{aligned}
      \mathbf{S}_0 &= E^2 = I \, , \\
      \mathbf{S}_1 &= I \cos 2\alpha \, , \\
      \mathbf{S}_2 &= I \sin 2\alpha \cos \beta \, , \\
      \mathbf{S}_3 &= - I \sin 2\alpha \sin \beta \, . \phantom{ZZZ}
   \end{aligned}
   \right\}
\end{equation}
Here, $2\alpha$ and $\beta$ are the polar and azimuthal angles, respectively, in a spherical coordinate system with $\mathbf{S}_1$ being the polar axis, and $\{ \mathbf{S}_2, \mathbf{S}_3\}$ the equatorial plane. Hence, if $\alpha = [0,\pi/2]$, and $\beta = [0,2\pi]$, we can cover the entire surface of the sphere. 

From \eqref{eq:stokesvector_alphabeta}, we see that the Stokes vector of $\hat{\mathbf{e}}_{A}$ (shown in \figref{fig:1} as $\mathbf{A}$) is equal to $\mathbf{S}_1$. Therefore, $2\alpha$ is the angle in Poincar\'e space that the SOP of $\mathbf{E}_C$ makes with the basis vector $\hat{\mathbf{e}}_A$ (cf.\ \figref{fig:1}). Moreover, if we can define $\angle AC = 2\alpha$, we can use \eqref{eq:Ec} to conclude that the intensity of the $\hat{\mathbf{e}}_A$-component is $I \cos^2(\angle AC / 2)$ and the intensity of the $\hat{\mathbf{e}}_{A'}$-component is $I \sin^2(\angle AC / 2)$, which is Pancharatnam's Theorem~1.

Pancharatnam next considers the problem of explaining interference between two non-orthogonal states of polarization. It was already well known that the interference of two beams of the same state of polarization (SOP) results in an intensity
\begin{equation}\label{eq:1}
    I (I_1,I_2,\delta) = I_1 + I_2 + 2 \sqrt{I_1 I_2} \cos(\delta) \, ,
\end{equation}
where $I_1$ and $I_2$ are the individual intensities, and $\delta$ is the phase difference between the two beams. We can see that changing $I_1$ or $I_2$ only modifies the modulation contrast (visibility), but not the phase of the sum wave $I$.\cite{Garza-Soto2020} 

In order to see what happens when two beams of different SOP are combined, Pancharatnam considers the superposition of beams in states $\mathbf{A}$ and $\mathbf{B}$. He considers one of them, e.g., $\mathbf{A}$, and its orthogonal $\mathbf{A}'$, as the polarization basis, and finds the intensity of the sum by applying the following steps:
\begin{enumerate}
    \item Coherently adding $\mathbf{A}$ to the $\mathbf{A}$-component of $\mathbf{B}$,
    \item Incoherently adding the $\mathbf{A}'$-component of $\mathbf{B}$ to the intensity obtained in step~1. 
\end{enumerate}
That is, the projection of $\mathbf{A}$ onto $\mathbf{B}$ gives the component of $\mathbf{A}$ that directly interferes with $\mathbf{A}$. The component of $\mathbf{A}$ orthogonal to $\mathbf{B}$ does not interfere, and so only adds a bias to the intensity.

Considering Pancharatnam's Theorem~1, we therefore decompose $\mathbf{B}$ into two beams of orthogonal polarizations, yielding the intensities
\begin{align}
    I_{B,A} = I_B \cos^2 (c/2) \, , \\
    I_{B,A'} = I_B \sin^2 (c/2) \, ,
\end{align}
where $I_B$ is the intensity of $\mathbf{B}$, and $c$ is the angle between the two states on the Poincar{\'e} sphere, and is thus twice the angle between the states in Cartesian space. For step 1, we coherently combine $I_A$ (the intensity of $\mathbf{A}$) with $I_{B,A}$ using \eqref{eq:1}, giving
\begin{align}
     &I (I_A,I_{B,A},\delta) = \notag \\
      &\quad I_A + I_B \cos^2(c/2) + 2 \big[ I_A I_B \cos^2(c/2) \big]^{1/2} \cos(\delta) \, .
\end{align}
This expression represents the intensity that results from the addition of the $\mathbf{A}$-component of both beams. 

Finally, with step~2, we add $I_{B,A'}$ to obtain the intensity
\begin{align}
    I &= I_A + I_B \cos^2(c/2) + I_B \sin^2(c/2) \notag \\
      &\quad\qquad + 2 \big[ I_A I_B \cos^2(c/2) \big]^{1/2} \cos(\delta) \, ,
\end{align}
which simplifies to
\begin{equation}\label{eq:Pancharatnam1}
    I = I_A + I_B + 2 \sqrt{I_A I_B} \cos(c/2) \cos(\delta) .
\end{equation}
The above equation is Pancharatnam's Eq.~(1) in Ref.~\citenum{Pancharatnam1956}. This expression has the advantage that the factor containing the phase $\delta$ between the beams and $\cos(c)$ --- what Pancharatnam calls their ``similarity factor'' --- are separated as a product.

\begin{figure}
    \centering
    \includegraphics[width=0.8\linewidth]{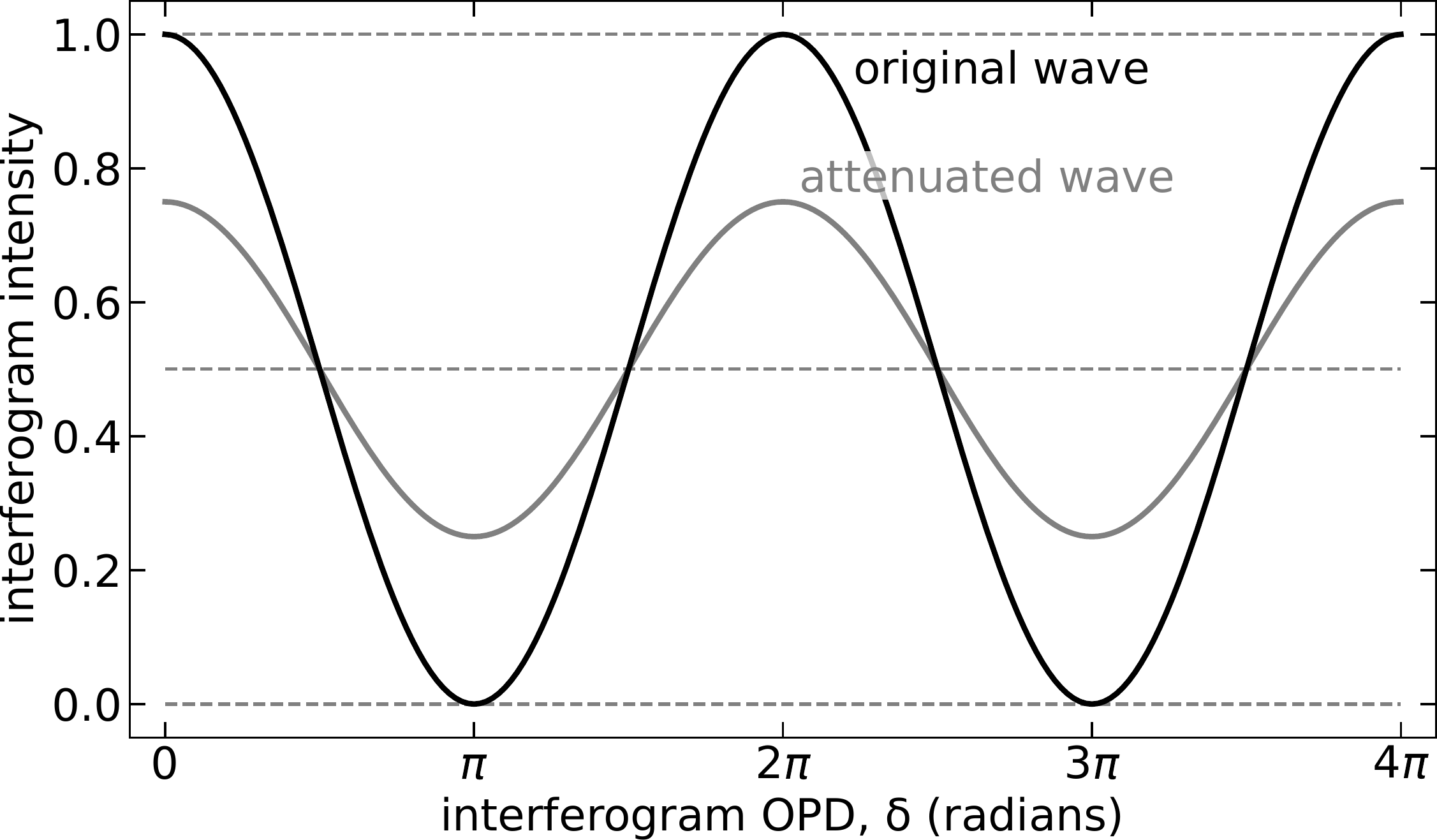}
    \caption{The ``original wave'' interferogram is obtained in \eqref{eq:Pancharatnam1} when the two intensities $I_A$ and $I_B$ are equal and held fixed as we vary the relative phase $\delta$ between them. The ``attenuated wave'' occurs when we reduce the intensity of one beam, and again vary the phase between the two beams.}
    \label{fig:attenuated_wave}
\end{figure}

Equation~\ref{eq:Pancharatnam1} is a useful result because it makes two properties clear. First, while $\delta$ is the phase delay between input beams $\mathbf{A}$ and $\mathbf{B}$, the equation shows that the intensity of the beams' superposition varies sinusoidally, with $\delta$ as the phase of $I$, independent of what polarization states are involved. This is illustrated in \figref{fig:attenuated_wave}. Second, if either of the two intensities $I_A$ or $I_B$ changes, this changes $I$ but does not change the value of $\delta$. Here we see that $c$ determines the fringe visibility of the interferogram. Therefore, this one of the results that Pancharatnam has in mind when he titles his paper the ``generalized theory of interference''.

To compare with more modern approaches, we can analyze the same case using the Jones calculus. In order to simplify the math, we will simply define the state of the first beam to be horizontal, while that of the second beam is oriented at an angle of $\theta$ with respect to the horizontal. We also allow there to be a propagation phase delay $\delta$ between the two beams. Therefore, we have the two electric field vectors 
\begin{equation}
\vv{E}_A = |\vv{E}_A| \hat{x}, \qquad \text{and} \qquad \vv{E}_B = |\vv{E}_B| e^{-i \delta} \left(\cos{\theta}\hat{x} + \sin{\theta}\hat{y}\right).
\end{equation}
The amplitude of their sum is
\begin{equation}
\vv{E}_A + \vv{E}_B = \left( |\vv{E}_A| + |\vv{E}_B| \cos \theta \, e^{-i \delta} \right)\hat{x} + |\vv{E}_B| \sin \theta \, e^{-i \delta}\hat{y},
\end{equation}

so that intensity of the two beams' interference is
\begin{align}
   I &= \big( \vv{E}_A + \vv{E}_B \big) \cdot \big( \vv{E}_A + \vv{E}_B \big)^{\ast} \notag \\
      &= \big( |\vv{E}_A| + |\vv{E}_B| \cos \theta \, e^{+i \delta} \big) \big( |\vv{E}_A| + |\vv{E}_B| \cos \theta \, e^{-i \delta} \big) \notag \\
         &\qquad + \big( |\vv{E}_B| \sin \theta \, e^{+i \delta} \big) \big( |\vv{E}_B| \sin \theta \, e^{-i \delta} \big) \notag \\
      &= |\vv{E}_A|^2 + |\vv{E}_B|^2 + 2 |\vv{E}_A| |\vv{E}_B| \cos \theta \cos \delta \, , \label{eq:jones}
\end{align}
which is exactly Pancharatnam's expression [given as \eqref{eq:Pancharatnam1} above], since $c = 2 \theta$.

\begin{figure*}
   \centering
   \includegraphics[width=0.85\linewidth]{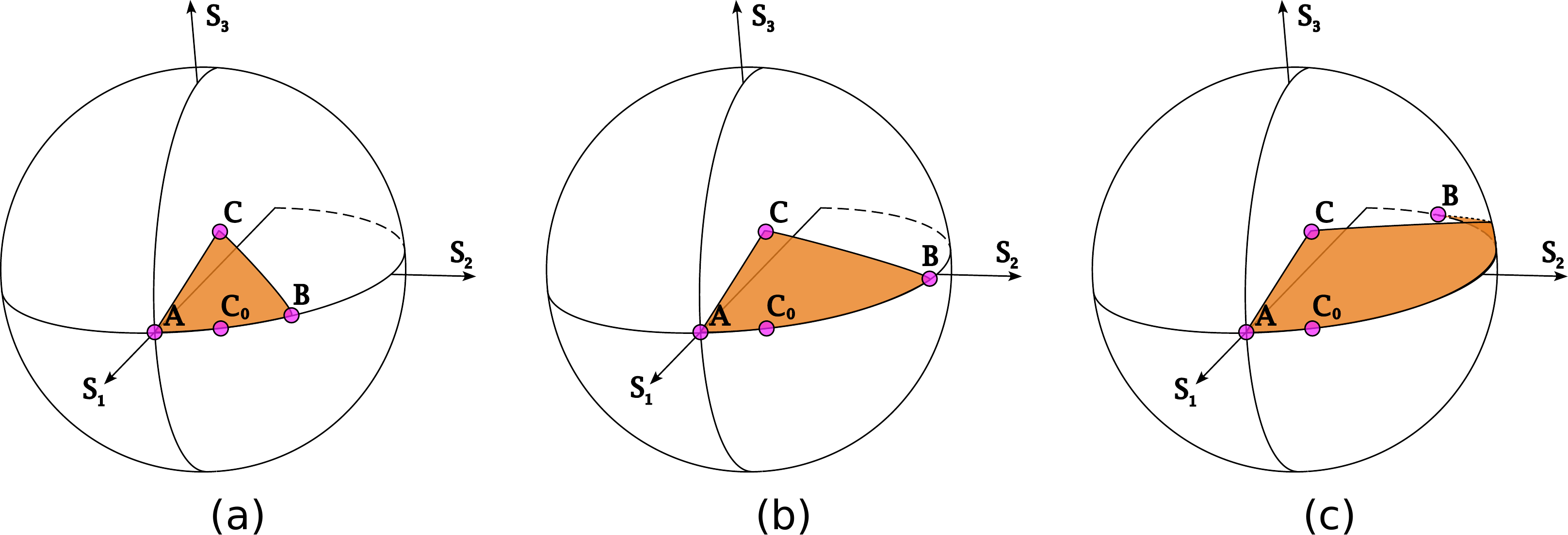}
   \caption{The increasing area (drawn in orange) on the surface of the sphere, formed by the spherical triangle between points $\mathbf{A}$, $\mathbf{B}$, and $\mathbf{C}$.}
   \label{fig:three_spheres}
\end{figure*}

This can be generalized to an arbitrary pair of elliptical states, considering two waves with electric field amplitudes $\mathbf{E}_A$ and $\mathbf{E}_B$, and a phase difference $\delta$. The intensity of their superposition is
\begin{align}
   I &= (\mathbf{E}_A + e^{i\delta} \mathbf{E}_B) \cdot (\mathbf{E}_A^{\ast} + e^{-i\delta} \mathbf{E}_B^{\ast}) \nonumber \\
     &= |\mathbf{E}_A|^2 + |\mathbf{E}_B|^2 + 2 \mathrm{Re} \left\{ e^{-i\delta}  (\mathbf{E}_A \cdot \mathbf{E}_B^{\ast}) \right\}.
\end{align}
The result of $\mathbf{E}_A \cdot \mathbf{E}_B^\ast$ can be inferred from the discussion about Pancharatnam's Theorem~1, which gives $E_A E_B \cos(c/2)$, where $E_A,E_B$ are the real amplitudes of the electric fields. Therefore, we conclude that   
\begin{align}
   I &= |\mathbf{E}_A|^2 + |\mathbf{E}_B|^2 + 2 E_A E_B \cos (c/2) \cos(\delta) \nonumber \\
     &= I_A + I_B + 2 \sqrt{I_A I_B} \cos (c/2) \cos(\delta),
\end{align}
which agrees with Pancharatnam's result.

\section{The relationship between intensities and locations on the Poincar{\'e} sphere}\label{sec:solid_angle}

Following his expression for the interference of nonorthogonal beams, Pancharatnam requires only a few short steps (Sec.~4 of his paper) to develop his famous solid angle formula for the geometric phase. Here he considers the decomposition of a beam of state of polarization $\mathbf{C}$ into two nonorthogonal beams of polarization states $\mathbf{A}$ and $\mathbf{B}$. If \eqref{eq:Pancharatnam1} is a general equation for the interference of two beams with different SOPs, then one should be able to re-arrange the equation to solve for the phase delay $\delta$ between them via the intensities of $\mathbf{A}$ and $\mathbf{B}$ (but which Pancharatnam labels $I_1$ and $I_2$):
\begin{equation}\label{eq:cos_delta}
   \cos (\delta) = \frac{I - (I_A + I_B)}{2 \sqrt{I_A I_B} \cos(c / 2)} \, .
\end{equation}
We can interpret this result as saying that ``whatever this phase $\delta$ may be, it maintains a specific relationship between the intensities of the beam being decomposed ($\mathbf{C}$) and the intensities of the beams that result from the decomposition ($\mathbf{A}$ and $\mathbf{B}$).''

While he could have finished with this numerical formula for $\delta$, he went a step further and realized that this equation expresses a solid angle relationship between the SOPs of the three beams. To explain this he uses electric field vectors $\vv{E}_A$ and $\vv{E}_B$ as components of a sum vector $\vv{E}_C = \vv{E}_A + \vv{E}_B$. From vector analysis, it is easy to see that the part of $\vv{E}_B$ perpendicular to $\vv{E}_A$ also has to be equal to the part of $\vv{E}_C$ perpendicular to $\vv{E}_A$. Writing this in terms of the angles $a$, $b$, and $c$ from \eqref{eq:abc}, we have
\begin{equation}
   \sqrt{I_B} \sin (c/2) = \sqrt{I} \sin (b/2) \, .
\end{equation}
Taking the square of the components to find the intensities results in the proportions
\begin{equation}
   I_A = I \frac{\sin^2 (a/2)}{\sin^2 (c/2)} \qquad \text{and} \qquad I_B = I \frac{\sin^2 (b/2)}{\sin^2 (c/2)} \, .
\end{equation}
By expressing the intensities $I_A$ and $I_B$ in terms of the total intensity $I$, together with the angles between states on the Poincar{\'e} sphere, we can rewrite \eqref{eq:cos_delta} as
\begin{equation}\label{eq:Pancharatnam4}
   \cos (\delta) = \frac{\sin^2 (c/2) - \sin^2 (a/2) - \sin^2 (b/2)}{2 \sin (a/2) \sin (b/2) \cos (c/2)} \, ,
\end{equation}
which is Pancharatnam's Eq.~4. Now the equation is entirely in terms of the angles between states, rather than their intensities, and he recognizes that the form of this expression is almost the same as that of a solid angle formula. In fact, if we re-express the above relationships in terms of states $\mathbf{A}$, $\mathbf{B}$, and $\mathbf{C'}$ rather than $\mathbf{A}$, $\mathbf{B}$, and $\mathbf{C}$, where $\mathbf{C'}$ is the antipodal point to $\mathbf{C}$ on the Poincar{\'e} sphere, we obtain
\begin{equation}
   \cos (\delta) = \frac{1 - \cos^2 (c/2) - \cos^2 (a' / 2) - \cos^2 (b' / 2)}{2 \cos (a' / 2) \cos (b' / 2) \cos (c/2)} \, ,
\end{equation}
which does have the recognizable form of a solid angle formula. The primes indicate that the angles are to be taken with respect to $\mathbf{C'}$ rather than $\mathbf{C}$.
That is, $b'/2 = \angle AC'$, and $a'/2 = \angle BC'$. Making use of the solid angle formula,\cite{todhunter1863spherical}
\begin{equation}\label{eq:omega}
   \delta = \Omega' / 2
\end{equation}
when $\Omega'$ is the angle subtended by the spherical triangle $\mathbf{A},\mathbf{B},\mathbf{C'}$ from the center of the sphere. (Pancharatnam actually expresses this as $\delta = \pi - \tfrac{1}{2} E'$, where $E'$ is the spherical excess of the triangle.) The sign of $\delta$ given here corresponds to describing the sequence of states $\mathbf{A} \to \mathbf{B} \to \mathbf{C'}$ in a clockwise sense. If the direction of the sequence is reversed, then the sign flips.

These last statements, in which Pancharatnam works out the correct sign of the solid angle, is the only location in the paper where he talks about a sequence of states. However, it is clear from context that he is not referring to a cycle of polarization states but rather to the phase relationships between the two output states $\mathbf{A}$ and $\mathbf{B}$, and the state $\mathbf{C}$ from which they were decomposed.

\section{Extending the reasoning to the orthogonal case}

Considering that Pancharatnam found the phase between two beams using intensity of interference, his definition does not apply for the case in which states $\mathbf{A}$ and $\mathbf{B}$ are orthogonal. However, he provides a geometric argument to show that the formula can still be applied in the limit as the states approach orthogonality. Figure~\ref{fig:three_spheres}a shows an initial situation with states of polarization $\mathbf{A}$ and $\mathbf{B}$, and the state $\mathbf{C}$ obtained by their sum. This is the therefore the inverse of the case treated in Sec.~\ref{sec:solid_angle} above, but which is described by \eqref{eq:Pancharatnam1} in Sec.~\ref{sec:start}. A point $\mathbf{C}_0$ lying on the geodesic arc $\mathbf{AB}$ describes a state of polarization that results from adding the beams of polarization $\mathbf{A}$ and $\mathbf{B}$ with no phase between them. Recalling that the solid angle used by Pancharatnam is the one subtended by triangle $\mathbf{A B C'}$ rather than $\mathbf{A B C}$, this situation gives a solid angle of $\Omega' = \pi$.

\begin{figure}[bht]
    \centering
    \includegraphics[width=0.65\linewidth]{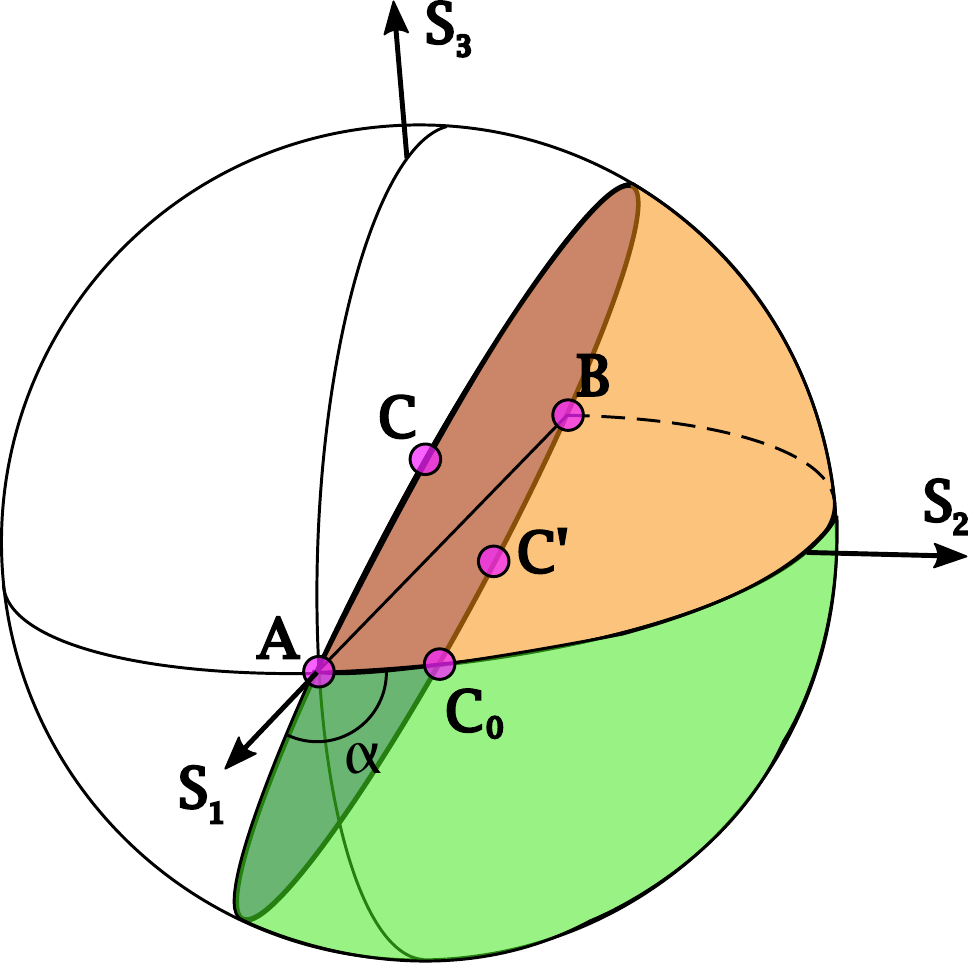}
    \caption{If we continue the evolution shown in \figref{fig:three_spheres} until state $\mathbf{B}$ becomes orthogonal to $\mathbf{A}$, then we form a spherical lune (drawn in orange) between geodesic arcs $\mathbf{A C_0 B}$ and $\mathbf{ACB}$.}
    \label{fig:sphere_orthogonal}
\end{figure}

Next we modify the polarization state of $\mathbf{B}$ so that it moves further away from $\mathbf{A}$ along the equator of the Poincar{\'e} sphere, as shown in \figref{fig:three_spheres}b, and then \figref{fig:three_spheres}c. As $\mathbf{B}$ moves away from $\mathbf{A}$, the enclosed solid angle subtended by $\mathbf{ABC}$ increases. As $\mathbf{B}$ approaches the point orthogonal to $\mathbf{A}$ (this point is labelled by Pancharatnam as $\mathbf{A'}$), the geodesic $\mathbf{ACB}$ becomes half a great circle, as in \figref{fig:sphere_orthogonal}. One might argue that in this case the enclosed area becomes undefined, since we can draw the geodesic connecting $\mathbf{A}$ and $\mathbf{B}$ in either a clockwise or an anticlockwise sense. However, if we note that the geodesic from $\mathbf{A}$ to $\mathbf{B}$ has until this point always passed through the intermediate point $\mathbf{C}_0$, then it will for this limit case as well. The solid angle that relates the phase between the two beams is therefore the one enclosed between the two geodesic arcs $\mathbf{A C_0 A'}$ and $\mathbf{A C' A'}$ (where $\mathbf{A'}$ coincides with the point written as $\mathbf{B}$ in \figref{fig:sphere_orthogonal}). This is a spherical lune (drawn in green in the figure) whose solid angle subtended from the origin is exactly twice the value of the angle $\alpha$ formed between states $\mathbf{C}_0$, $\mathbf{A}$, and $\mathbf{C'}$ at the surface of the sphere.

Using \eqref{eq:omega} finally allows us to equate the phase difference between beams $\mathbf{A}$ and $\mathbf{B}$ to the angle $\alpha$: $\delta = \alpha$. That is, the angle $\alpha$ (or half the solid angle $\Omega$) is equal to the phase $\delta$ that one must retard state $\mathbf{A'}$ from $\mathbf{A}$ in order that both be correctly decomposed from input state $\mathbf{C}$.

\section{Interference of the components transmitted by an analyzer}

It was of practical importance for Pancharatnam to have an expression for the phase of two arbitrarily-polarized beams transmitted through an analyzer because this was the configuration has was using to measure the pattern transmitted by biaxial crystals such as iolite. Finding such an expression is the purpose of Section~8 of Pancharatnam's manuscript.

\begin{figure}[bht]
    \centering    
    \includegraphics[width=0.65\linewidth]{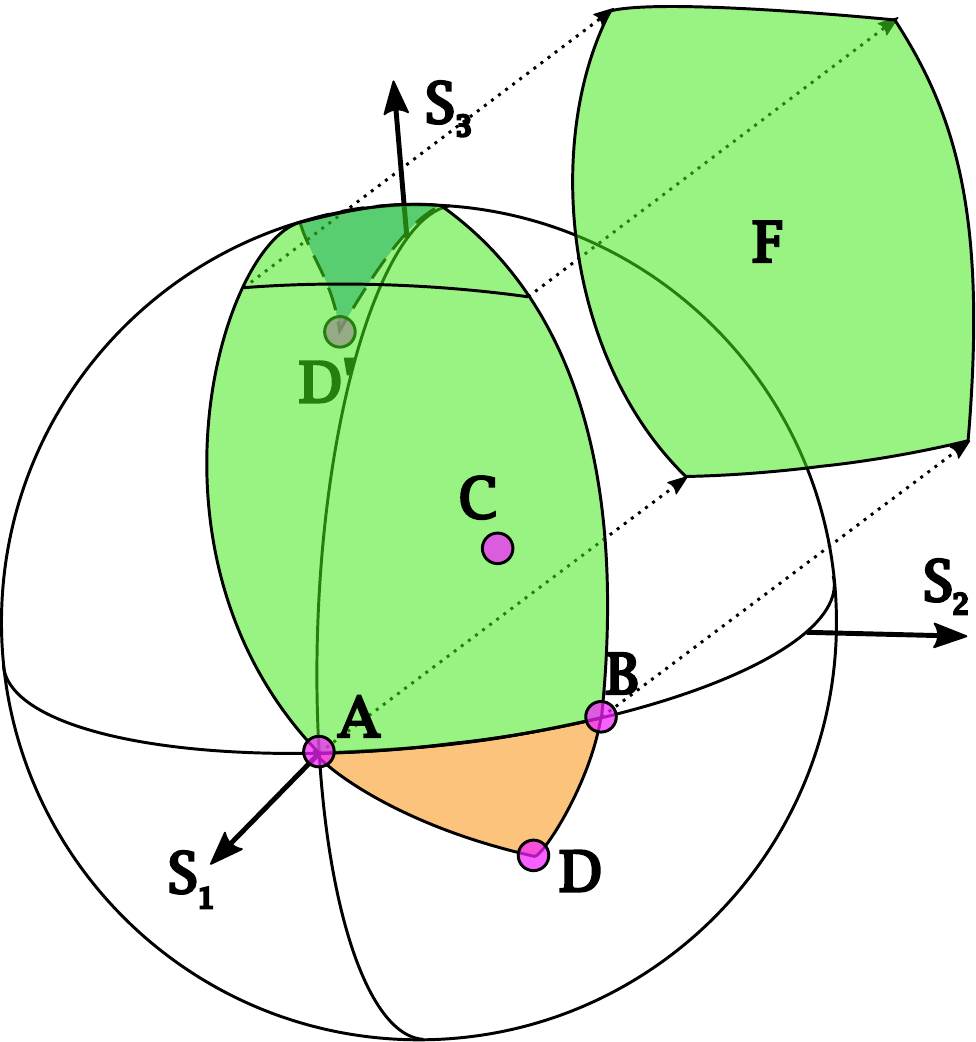}
    \caption{The spherical surface elements used to calculate the intensity transmitted by an arbitrary analyzer (state $\mathbf{D}$) when two states $\mathbf{A}$ and $\mathbf{B}$ are incident upon it.}
    \label{fig:sphere_abcprime}
\end{figure}

Pancharatnam first defines $\mathbf{D}$ as the state of polarization transmitted by the analyzer. Unfortunately, Pancharatnam once again reuses the symbol $\mathbf{C}$ here. We write this as $\mathbf{D}$ to avoid confusion with the states defined in the earlier sections. He also uses $I_1$ and $I_2$ in place of $I_A$ and $I_B$, but we have retained the latter for consistency. Pancharatnam also reuses the symbols $a$, $b$, and $c$ for the angles between states, but since the definition of $\mathbf{C}$ has changed, we instead define $\theta_{ij}$ to represent the angle between points $i$ and $j$ on the surface of the Poincar{\'e} sphere, as subtended from the center of the sphere. His goal is to relate the phase of the interference transmitted by the analyzer to the phase difference $\delta$ between the two input states (see \figref{fig:sphere_abcprime}).

When beams $\mathbf{A}$ and $\mathbf{B}$ are incident on the analyzer, the transmitted intensity $I_D$ will be the component of state $\mathbf{A}$ along direction $\mathbf{D}$, plus the component of state $\mathbf{B}$ also along $\mathbf{D}$, while incorporating an unknown phase difference $\delta'$ between $\mathbf{A}$ and $\mathbf{B}$:
\begin{align}
   I_D &= I_A \cos^2 (\theta_{AD} / 2) + I_B \cos^2 (\theta_{BD} / 2) \notag \\
      &\qquad+ \sqrt{I_A I_B} \cos (\theta_{AD} / 2) \cos (\theta_{BD} / 2) \cos (\delta') \, . \label{eq:Id}
\end{align}

If we use the analyzer oriented at angle $\mathbf{D'}$ orthogonal to $\mathbf{D}$, then we would get a different intensity $I_{D'}$ and an unknown phase difference $\delta''$:
\begin{align}
   I_{D'} &= I_A \sin^2 (\theta_{AD'} / 2) + I_B \sin^2 (\theta_{BD'} / 2) \notag \\
      &\qquad+ \sqrt{I_A I_B} \sin (\theta_{AD'} / 2) \sin (\theta_{BD'} / 2) \cos (\delta'') \, .
\end{align}
From \figref{fig:sphere_abcprime}, we can see that the area enclosed by $\mathbf{D A D' B D}$ is a spherical lune. The phase $\delta'$ needed to generate state $\mathbf{D}$ from adding $\mathbf{A}$ and $\mathbf{B}$ is given by $\delta' = \pi - \tfrac{1}{2} E'$, where $E'$ is the solid angle subtended by the spherical triangle $\mathbf{ABD'}$ (drawn in green in the figure). In a similar fashion, the phase $\delta''$ needed when adding $\mathbf{A}$ and $\mathbf{B}$ to get $\mathbf{D'}$ is given by $\delta'' = \pi - \tfrac{1}{2} E''$, where $E''$ is the solid angle of $\mathbf{ABD}$ (drawn in orange).

Since both $\delta'$ and $\delta''$ are equivalent to corresponding solid angles, if we subtract the two, we obtain the solid angle subtended by quadrangular area $F$ in the sphere\cite{Gutierrez-Vega2011,Kurzynowski2011}
\begin{equation}
   F = \pm (\delta'' - \delta') \, .
\end{equation} 
Confusingly, Pancharatnam yet again reuses the symbol $C$ to indicate this quadrangular area. We will instead use $F$, as \figref{fig:sphere_abcprime} does. The sign of $F$ is determined by whether the sequence of points $\mathbf{ABD}$ proceeds in a clockwise or anticlockwise fashion. Pancharatnam then takes both $I_D$ and $I_{D'}$ and adds them together to get the total intensity, equal to the intensity of $\mathbf{C}$ that depends on $\delta$ but now expressed in terms of $\delta'$ and $F$:
\begin{align}
   I &= I_A + I_B + 2 \sqrt{I_A I_B} \big[ \cos (\theta_{AD} / 2) \cos (\theta_{BD} / 2) \cos (\delta') \notag \\
      &\hspace{2cm} + \sin (\theta_{AD'} / 2) \sin (\theta_{BD'} / 2) \cos (\delta' \pm F) \big] \, .
\end{align}
Now that we have the same angle $\delta'$ in both terms inside the square brackets, we can recognize that this is a standard expression for the spherical excess of a triangle, so that the equation simplifies to~\cite{todhunter1863spherical}
\begin{equation}\label{eq:last}
   I = I_A + I_B + 2 \sqrt{I_A I_B} \cos (c / 2) \cos (\delta' + \tfrac{1}{2} E) \, .
\end{equation} 
This equation is now identical in form to \eqref{eq:Pancharatnam1}, from which we can then say that $\delta' = \delta - \tfrac{1}{2} E$. Now we have a means of calculating $\delta'$ for a given pair of input states $\mathbf{A}$ and $\mathbf{B}$, together with the analyzer orientation giving $\mathbf{D}$. Using this in \eqref{eq:Id}, we can now calculate the intensity transmitted by the analyzer depending on the original phase difference $\delta$ between the two input beams.

\section{Conclusion}

Through an impressive set of spherical trigonometry manipulations on the Poincar{\'e} sphere, Pancharatnam found that polarization states have specific phase relationships that are generally not taken into account, unless one is performing interferometric measurements. This was the case for him, since he was analyzing the light transmitted through dichroic biaxial crystals. The correct analysis of these measurements required that he incorporate this new phase, what we now refer to as the geometric phase, into his equations.

Among his less-known results is Equation~\ref{eq:cos_delta}, which shows that if one knows the intensities of two input beams as well as the intensity of their interference, one can infer the phase difference $\delta$ between the two input beams from these simple intensity measurements alone. Also notable is \eqref{eq:last}, which demonstrates that the phase of transmitted by an analyzer is not in general equal to the phase of a wave incident upon it.

The recently published ``wave description of geometric phase'' interprets the geometric phase as a shift in the wave peak location away from the midpoint between the peaks of the two input waves.\cite{Garza-Soto2022} One can see the close connection between the wave description and Pancharatnam's approach by considering \eqref{eq:Pancharatnam1} (Pancharatnam's Eq.~1), which expresses the intensity produced by adding two waves as being a bias value ($I_1 + I_2$) plus a single cosine term with amplitude $\sqrt{I_1 I_2} \cos (c/2)$. Thus, Pancharatnam is also considering the case of two cosine waves summing together into a single cosine output wave.

In \eqref{eq:jones}, we were also able to show how Pancharatnam's results are derived using the familiar modern approach, not available to Pancharatnam, of the Jones calculus. The Jones calculus explicitly forms a pair of 2D electric field vectors, and adding these two waves produces a cosine factor $\cos (\delta)$ that includes the same phase delay that Pancharatnam obtained, but which does not seem to have been replicated until Michael Berry's work in 1987.\cite{Berry1987}

Berry naturally interpreted Pancharatnam's work through the lens of his own work, which approached the geometric phase as a shift resulting from the evolution of a quantum state around a cycle in parameter space. Upon reading Pancharatnam's approach, he saw that this could easily be a cycle of polarization states generating this phase shift. Pancharatnam, however, was focused on interferometric measurement and not on modeling the evolution of polarization states. As it turns out, the two are equivalent, and so the misunderstanding is not at all a serious one. In the literature, many authors actually refer to this shift due to polarization state evolution as a ``Pancharatnam-Berry phase''. This seems the ideal choice, since ``Pancharatnam phase'' should perhaps be more narrowly defined as a shift resulting from adding polarized waves.


\begin{thebibliography}{10}
   \newcommand{\enquote}[1]{``#1''}

   \bibitem{Ramaseshan1988}
   S.~Ramaseshan, \enquote{The {P}oincare sphere and the {P}ancharatnam phase ---
   some historical remarks,} {\textit{Current Science}}
   \textbf{59} (1990).

   \bibitem{Nityananda2013}
   R.~Nityananda, K.~Ramaseshan, N.~Madhusudana, and G.~Series, \enquote{S
   pancharatnam (1934--1969): three phases,} {\textit{Resonance}}
   \textbf{18}, 301--305 (2013).

   \bibitem{Pancharatnam1956}
   S.~Pancharatnam, \enquote{Generalized theory of interference and its
   applications, part {I}. {C}oherent pencils,}
   {\textit{Proceedings of the Indian Academy of
   Sciences---Section A}} \textbf{44}, 398--417 (1956).

   \bibitem{Berry1987}
   M.~V. Berry, \enquote{The adiabatic phase and {P}ancharatnam's phase for
   polarized light,} {\textit{J. Mod. Opt.}} \textbf{34},
   1401--1407 (1987).

   \bibitem{Aravind1992}
   P.~K. Aravind, \enquote{A simple proof of pancharatnam's theorem,}
   {\textit{Opt. Comm.}} \textbf{094}, 191--196 (1992).

   \bibitem{Tiwari1992}
   S.~C. Tiwari, \enquote{Geometric phase in optics: quantal or classical?}
   {\textit{J. Mod. Opt.}} \textbf{39}, 1097--1104 (1992).

   \bibitem{Roy2002}
   M.~Roy, P.~Svahn, L.~Cherel, and C.~J.~R. Sheppard, \enquote{Geometric
   phase-shifting for low-coherence interference microscopy,}
   {\textit{Optics and Lasers in Engineering}} \textbf{37},
   631--641 (2002).

   \bibitem{Kurzynowski2011}
   P.~Kurzynowski, W.~A. Woźniak, and M.~Szarycz, \enquote{Geometric phase: two
   triangles on the {P}oincar{\'e} sphere,} {\textit{J. Opt. Soc.
   Am. A}} \textbf{28}, 475--482 (2011).

   \bibitem{Lages2014}
   J.~Lages, R.~Giust, and J.-M. Vigoureux, \enquote{Geometric phase and
   {P}ancharatnam phase induced by light wave polarization,}
   {\textit{Physica E}} \textbf{59}, 6--14 (2014).

   \bibitem{Cohen2019}
   E.~Cohen, H.~Larocque, F.~Bouchard, F.~Nejadsattari, Y.~Gefen, and E.~Karimi,
   \enquote{Geometric phase from {A}haronov-{B}ohm to {P}ancharatnam-{B}erry and
   beyond,} {\textit{Nature Rev. Phys.}} \textbf{1}, 437--449
   (2019).

   \bibitem{Arteaga2021}
   O.~Arteaga, \enquote{Fresnel-{A}rago fifth law of interference: the first
   description of a geometric phase in optics,} {\textit{J. Mod.
   Opt.}} \textbf{68}, 350--357 (2021).

   \bibitem{Berry1984}
   M.~V. Berry, \enquote{Quantal phase factors accompanying adiabatic changes,}
   {\textit{Proc. Roy. Soc. London A}} \textbf{392}, 45--54
   (1984).

   \bibitem{Simon1983}
   B.~Simon, \enquote{Holonomy, the quantum adiabatic theorem, and {B}erry's
   phase,} {\textit{Phys. Rev. Lett.}} \textbf{51}, 2167--2170
   (1983).

   \bibitem{Ramaseshan1986}
   S.~Ramaseshan and R.~Nityananda, \enquote{The interference of polarized light
   as an early example of {B}erry's phase,} {\textit{Current
   Science}} \textbf{55}, 1225--1226 (1986).

   \bibitem{Berry1994}
   M.~Berry, \enquote{Pancharatnam, virtuoso of the {P}oincar{\'e} sphere: an
   appreciation,} {\textit{Current Science}} \textbf{67}, 220--223
   (1994).

   \bibitem{Berry2010}
   M.~Berry, \enquote{Geometric phase memories,} {\textit{Nature
   Physics}} \textbf{6}, 148--150 (2010).

   \bibitem{Vinitskii1990}
   S.~I. Vinitski{\u\i}, V.~L. Derbov, V.~M. Dubovik, B.~L. Markovski, and Y.~P.
   Stepanovski{\u\i}, \enquote{Topological phases in quantum mechanics and
   polarization optics,} {\textit{Soviet Physics Uspekhi}}
   \textbf{33}, 403 (1990).

   \bibitem{Jackson2008}
   J.~D. Jackson, \enquote{Examples of the zeroth theorem of the history of
   science,} {\textit{Am. J. Phys.}} \textbf{76}, 704--719 (2008).

   \bibitem{collett2005field}
   E.~Collett, \emph{{Field Guide to Polarization}}, Field Guide Series (Society
   of Photo Optical, 2005).

   \bibitem{Garza-Soto2020}
   L.~Garza-Soto, A.~De-Luna-Pamanes, I.~Melendez-Montoya, N.~Sanchez-Soria,
   D.~Gonzalez-Hernandez, and D.~Lopez-Mago, \enquote{Geometric-phase
   polarimetry,} {\textit{J. Optics}} \textbf{22}, 125606--125615
   (2020).

   \bibitem{todhunter1863spherical}
   I.~Todhunter, \emph{{Spherical Trigonometry: For the Use of Colleges and
   Schools, with Numerous Examples}} (Macmillan, 1863).

   \bibitem{Gutierrez-Vega2011}
   J.~C. Gutiérrez-Vega, \enquote{Pancharatnam-{B}erry phase of optical systems,}
   {\textit{Opt. Lett.}} \textbf{36}, 1143--1145 (2011).

   \bibitem{Garza-Soto2022}
   L.~Garza-Soto, N.~Hagen, D.~Lopez-Mago, and Y.~Otani, \enquote{Wave description
   of geometric phase,} Submitted to JOSA-A, 2022.
\end{thebibliography}
\end{document}